\newcommand{\PRL}[3]{Phys.\ Rev.\ Lett.\ {\bf #1},\ #2 (#3)}
\newcommand{\RMP}[3]{Rev.\ Mod.\ Phys.\ {\bf #1},\ #2 (#3)}

\newcommand{\OC}[3]{Opt.\ Commun.\ {\bf #1},\ #2 (#3)}

\newcommand{\NAT}[3]{Nature\ {\bf #1},\ #2 (#3)}

\newcommand{\PRA}[3]{Phys.\ Rev.\ A\ {\bf #1},\ #2 (#3)}

\newcommand{\JPA}[3]{J.\ Phys.\ A:\ Math.\ Gen.\ {\bf #1},\ #2 (#3)}

\newcommand{\NJP}[3]{New.\ Jour.\ Phys.\ {\bf #1},\ #2 (#3)}

\newcommand{\APPA}[3]{Acta.\ Phys.\ Polon.\ A\ {\bf #1},\ #2 (#3)}

\newcommand{\IJQI}[3]{Int.\ jour.\ Quantum. \ info.\  {\bf #1},\ #2 (#3)}

\documentclass[aps,showkeys,showpacs,amsmath,amssymb]{revtex4}
\usepackage{dcolumn}
\usepackage{amsmath}
\usepackage{longtable}
\input{Qcircuit}
\begin{document}
\title{Secure quantum conversation through non-destructive discrimination of highly entangled multipartite states}

\author{Sakshi Jain}
\affiliation{Indian Institute of Technology-Bombay, Mumbai - 400076, India}
\email{sakshi.r.jain@gmail.com}
\author{Sreraman Muralidharan}
\affiliation{ Department of Information and communication technology, \\Royal Institute of Technology(KTH), SE-164 40 Kista, Sweden}
\email{sreraman@kth.se}
\author{Prasanta K. Panigrahi}
\email{prasanta@prl.res.in}
\affiliation{Indian Institute of Science Education and Research Kolkata, Mohanpur Campus, BCKV Campus Main Office, Mohanpur - 741252, India}
\affiliation{Physical Research Laboratory, Navrangpura, Ahmedabad - 380 009, India}

\begin{abstract}
 ``Quantum conversation'' is a way in which two parties can communicate classical information with each other using entanglement as a shared resource. We present this scheme using a multipartite entangled state after describing its generation through appropriate circuit diagrams. We make use of a discrimination scheme which allows one to perform a measurement on the system without destroying its entanglement. We later prove that this scheme is secure in a noiseless and a lossless quantum channel.   

\end{abstract}
\maketitle

\section{Introduction}
Quantum cryptography \cite{QCB, Hillery, Ekert} is one of the significant applications of quantum information theory, which has also been experimentally realized in various systems \cite{exp1, exp2}. It enables one to transmit secret messages between various parties with unconditional security. Quantum secure direct communication (QSDC) \cite{qsdc} is an important branch of quantum cryptography, which uses entanglement as a resource to transmit quantum information in a  secure manner, without establishing a random key. In recent years, several QSDC schemes, involving EPR pairs \cite{epr} and multiparticle entangled channels \cite{multi1, multi2} have been investigated. Most of these protocols involve joint measurements to be carried out on an entangled basis which is a non-trivial task. Instead, measurements on a product basis is often preferred \cite{pro}.  However, this disturbs the entanglement of the system and makes the state unusable for further applications. Instead, approaches based on non-destructive discrimination (NDD) can be more useful as it enables discrimination between orthogonal quantum states by allowing a product basis measurement without destroying the entanglement of the system \cite{Sakshi, Grassl1, Grassl2, Manu1}.

 Characterization of entanglement of greater than four qubits under LOCC is still an unsolved problem. However, approaches based on numerical optimization procedures to find higher qubit entangled states have recently attracted attention \cite{Brown, Borras, Stepney}. First such procedure was
 studied by Brown \textit{et al}. \cite{Brown}, who arrived at a ``highly entangled'' five qubit state given by, 
\begin{equation}
|\psi_{5}\rangle=\frac{1}{2}(|001\rangle|\phi_{-}\rangle+|010\rangle|\psi_{-}\rangle
+|100\rangle|\phi_{+}\rangle+|111\rangle|\psi_{+}\rangle),
\end{equation}
where $|\psi_{\pm}\rangle = \frac{1}{\sqrt{2}}(|00\rangle$ $\pm$ $|11\rangle)$, $|\phi_{\pm}\rangle = \frac{1}{\sqrt{2}}(|01\rangle$ $\pm$ $|10\rangle)$ are the Bell states. This result was verified by yet another numerical procedure carried out later \cite{Borras}.  This state is maximally ``persistent'' \cite{Hans} and shows a high degree of ``connectedness'' \cite{Hans2}, which is an indication of genuine multiparticle entanglement. Further, it assumes the same form for all ten $(3 + 2)$ splits. It was shown that, $|\psi_{5}\rangle$ can be used for perfect teleportation and state sharing of an arbitrary two qubit state \cite{Sre1}. This state can also be physically realized in a cavity QED system, which is insensitive to both cavity decay and thermal field \cite{qed}. The dense coding \cite{dense} capacity of this state reaches the ``Holevo bound'', thereby allowing the  transmission of five classical bits using  three qubits by consuming only two ebits of entanglement \cite{Sre1}. In this paper, we use the dense coding protocol and an efficient NDD scheme to set up a quantum conversation between two parties. We later prove that this protocol is secure under ideal conditions.

 For the sake of completeness, we shall discuss the dense coding protocol using $|\psi_{5}\rangle$ as a shared entangled resource \cite{Sre1}. Initially Alice and Bob possess, the first three and the last two qubits of $|\psi_{5}\rangle$ respectively. Suppose Alice has a secret message
 ``$m_{1}m_{2}m_{3}m_{4}m_{5}$''  where $m_{i}$ $\in$ \{0, 1\} which she wants to send to Bob. To achieve this purpose, Alice operates unitarily on her first three qubits as,
\begin{equation}
\mathbf{
((\zeta^{1} . \zeta^{2} . \zeta^{3} . \zeta^{4} . \zeta^{5}) \otimes I \otimes I)|\psi_5\rangle = |\psi_{5}\rangle_{m_{1}m_{2}m_{3}m_{4}m_{5}}}.
\end{equation}

\begin{eqnarray}
\zeta^{1} &=& \sigma_{x} \otimes I \otimes I, \text{if} \ { m_{1} = 1,} \\ \nonumber
          &=& I \otimes I \otimes I, \text{if} \  m_{1} = 0, \\ \nonumber
\zeta^{2} &=& I \otimes I \otimes \sigma_{x},  \text{if} \  m_{2} = 1, \\ \nonumber
          &=& I \otimes I \otimes I, \text{if} \ m_{2} = 0, \\ \nonumber
\zeta^{3} &=& \sigma_{z} \otimes \sigma_{z} \otimes I, \text{if} \ m_{3} = 1, \\ \nonumber
          &=& I \otimes I \otimes I, \text{if} \ m_{3} = 0, \\ \nonumber 
\zeta^{4} &=& \sigma_{x} \otimes I \otimes \sigma_{x}, \text{if} \  m_{4} = 1, \\ \nonumber
          &=& I \otimes I \otimes I,\text{if} \  m_{4} = 0, \\ \nonumber 
\zeta^{5} &=& I \otimes \sigma_{x} \otimes \sigma_{x}, \text{if} \  m_{5} = 1, \\ \nonumber
          &=& I \otimes I \otimes I, \text{if} \ m_{5} = 0.
\end{eqnarray}
 For example, if Alice wants to send the 5 bit classical information "10010" to Bob, the unitary operation that Alice would apply on her qubits would be;
 \begin{equation}
\mathbf{ (((\sigma_{x} \otimes I \otimes I)(I \otimes I \otimes I)(I \otimes I \otimes I)(\sigma_{x} \otimes I \otimes \sigma_{x})(I \otimes I \otimes I)) \otimes I \otimes I) |\psi_5\rangle = (|\psi_5\rangle_\mathbf{{100010}})}
 \end{equation}
  After performing the unitary operations in the above stated manner, Alice sends the three qubits to Bob.\normalfont Bob, on receiving Alice's qubits, performs a joint five partite von-Neumann measurement in the $\mathbf{|\psi_{5}\rangle_{m_{1}m_{2}m_{3}m_{4}m_{5}}}$ basis and distinguishes these
 states, thereby obtaining the message encoded by Alice.

The present scheme for bidirectional ``quantum conversation'' consists of five main steps. In the first and the second step, we generate the entangled state and initialize the protocol for a security check. In the third step, an eavesdropper's check on the multiqubit quantum channel is performed. Later, we propose a scheme to discriminate between the 32 orthogonal $\mathbf{|\psi_{5}\rangle_{m_{1}m_{2}m_{3}m_{4}m_{5}}}$ states after implementing the dense coding protocol thereby allowing the quantum conversation to be established. Subsequently, we check for the security of the presented protocol.
\section{Quantum Conversation}
\subsection{Generation and Initialization}
Here, we describe an explicit circuit diagram for the generation of the Brown \textit{et al}. state - a state which was obtained only through numerical searches in the past. 
 \begin{figure}[h]
	\caption{Circuit diagram for the generation of $|\psi_{5}\rangle$}
	\label{fig:CircuitDiagram}
	\leavevmode
\centering	
\Qcircuit @C=3.0em @R=2.0em {
\lstick{\ket{0}_{1}}& \gate{H} & \ctrl{1} &\qw &\ctrl{1}&\qw &\qw &\qw &\qw \\
\lstick{\ket{0}_{2}}& \gate{\sigma_x}      & \targ &\targ &\qw\qwx &\ctrl{1}&\qw &\qw &\qw\\
\lstick{\ket{0}_{3}}& \gate{H} &\qw &\ctrl{-1}&\qw\qwx &\qw\qwx &\qw &\qw  &\qw &|\psi_{5}\rangle\\
\lstick{\ket{0}_{4}}& \gate{\sigma_x}  &\qw &\qw &\targ\qwx &\qw\qwx &\gate{H} &\ctrl{1} &\qw\\
\lstick{\ket{0}_{5}}& \gate{\sigma_x}  &\qw &\qw &\qw &\targ\qwx &\qw &\targ\qwx &\qw
\ \gategroup{1}{9}{5}{9}{.7em}{\}} }  \\
\end{figure}
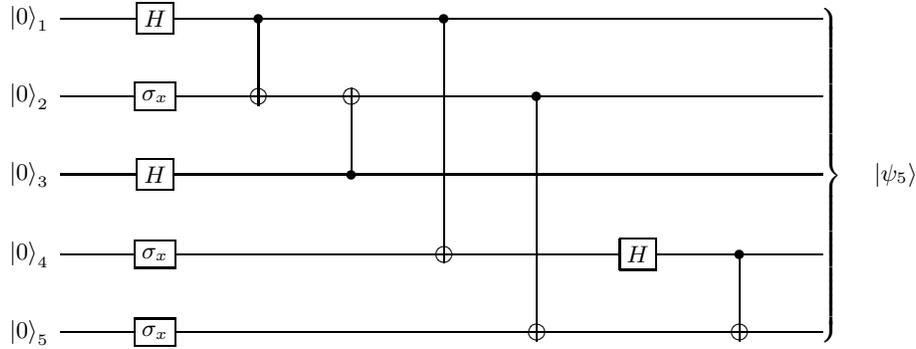
As is evident from Fig.1, by changing the input $|00000\rangle$ to a different computational basis, 32 orthogonal Brown \textit{et al}., can be obtained. The above circuit makes the Brown \textit{et al.}, state experimentally realizable in NMR and quantum dot systems.

In our quantum conversation protocol, Alice first prepares an ordered
sequence of $N$ copies of the five qubit Brown \textit{et al}., state
$|\psi_{5}\rangle$: [($q_{1}^{1}$, $q_{2}^{1}$, $q_{3}^{1}$,
  $q_{4}^{1}$, $q_{5}^{1})$), ($q_{1}^{2}$, $q_{2}^{2}$,......,
  $q_{5}^{N})$ ]. Here the subscripts represent five different qubits
in $|\psi_{5}\rangle$ and the superscripts indicate sequential
ordering of the $N$ different Brown \textit{et al.}, states.  Alice
then takes the same one qubit from each $|\psi_{5}\rangle$ to form
five ordered sequences corresponding to the five qubits, generically
written as,
\begin{equation}
S_{k}=[q_{k}^{1}, q_{k}^{2}, q_{k}^{3}.... q_{k}^{N}].
\end{equation}
Subsequently, she keeps the particle sequences $S_{1}$, $S_{2}$, $S_{3}$ and transmits the sequences $S_{4}$ and $S_{5}$ to Bob.

\subsection{Measurement basis of Alice and Bob}
Interestingly, the state $|\psi_{5}\rangle$ can be written down in two different forms as:
\begin{eqnarray}
|\psi_{5}\rangle_{12345}&=&\frac{1}{2}(|001\rangle|\phi_{-}\rangle+|010\rangle|\psi_{-}\rangle+|100\rangle|\phi_{+}\rangle+|111\rangle|\psi_{+}\rangle)_{12345},\\
|\psi_{5}\rangle_{13245}&=&  ([ |\phi_{+}\rangle|1\rangle + |\psi_{+}\rangle|0\rangle]|0+\rangle + [ |\phi_{+}\rangle|1\rangle -|\psi_{+}\rangle|0\rangle]|0-\rangle)_{13245} \\ \nonumber
& &- [ |\phi_{-}\rangle|1\rangle + |\psi_{-}\rangle|0\rangle]|1+\rangle + [ |\phi_{-}\rangle|1\rangle - |\psi_{-}\rangle|0\rangle]|1-\rangle.
\end{eqnarray}
Alice and Bob exploit this characteristic of $|\psi_{5}\rangle$ to check the presence of an eavesdropper. It can be noted that if one measures the qubits 4 and 5 by projection on the Bell basis, i.e., say $BMB_{1}$ (Bob's Measurement Basis-1), the first three qubits of Alice would collapse into one of the computational basis $(|001\rangle, |010\rangle, |100\rangle, |111\rangle)_{123}$

which now forms the corresponding basis of Alice's measurement ($AMB_{1}$). On the other hand, if the state $|\psi_{5}\rangle$ is written  as in equation (7), it can be observed that if Bob measures his two qubits in the basis ($BMB_{2}$): $|0+\rangle_{45}, |0-\rangle_{45}, |1+\rangle_{45}, |1-\rangle_{45}$, corresponding Alice's measurement basis ($AMB_{2}$) would be:

$(|\phi_{+}\rangle|1\rangle + |\psi_{+}\rangle|0\rangle)_{132}$, ($|\phi_{+}\rangle|1\rangle - |\psi_{+}\rangle|0\rangle)_{132}$, $(-|\phi_{-}\rangle|1\rangle$ - $|\psi_{-}\rangle|0\rangle)_{132}$ and ($|\phi_{-}\rangle|1\rangle - |\psi_{-}\rangle|0\rangle)_{132}$.  Utilizing this feature, Alice and Bob can design a defence strategy to guard against eavesdropping in the transmission. An explicit description of this strategy is given below.

\subsection{Eavesdropping check}
 Bob chooses randomly a sufficiently large subset (called the sample subset) of the pair of qubits 4 and 5 from the sequences $S_{4}$ and $S_{5}$
 and measures them either in the basis $BMB_{1}$ or $BMB_{2}$. He then tells Alice the position of these pairs and the corresponding chosen measurement
 basis through a classical channel. Alice then measures the set of qubits 1, 2, 3 from the sequences $S_{1}$, $S_{2}$, $S_{3}$ in the
 corresponding basis $AMB_{1}$ or $AMB_{2}$. Finally Alice and Bob check their measurement outcomes to find whether the quantum channel is altered.
 Later, they apply a majority voting technique to determine the error rate. If it turns out to be considerably high, they discard the sample subset and abort the protocol. Else, they will securely use the remaining entangled pairs of shared $|\psi_{5}\rangle$ to communicate their secret messages as in the next step.

\subsection{Dense coding and NDD}
The secret message ``$m_{1}m_{2}m_{3}m_{4}m_{5}$'' is encoded by Alice in her three qubits by applying the corresponding local unitary operations as in the dense coding protocol described earlier and sent to Bob. On receiving Alice's qubits, Bob applies non-destructive discrimination (NDD) on the five qubit Brown \textit{et al.}, state thus obtained and decodes the message without disturbing the entanglement of the system. Inorder to achieve this, he operates on the system through the circuit diagram shown in Fig.2, to discriminate between the 32 orthogonal Brown states non-destructively. \textbf{The outcome of the circuit with the input state as $\mathbf{|\psi_5\rangle_{m_{1}m_{2}m_{3}m_{4}m_{5}}}$ can be written in general as $\mathbf{|\psi_5\rangle_{x}|x\rangle}$ where $\mathbf{|x\rangle}$ is given by, $\mathbf{|m_1m_2m_3m_4m_5\rangle}$ and $\mathbf{|\psi_5\rangle_{x}}$ is given by, $\mathbf{|\psi\rangle_{m_{1}m_{2}m_{3}m_{4}m_{5}}}$.}  Thus, Bob now measures the ancillas in the product basis and thereby obtains the classical message.

 \begin{figure}[h]
	\caption{Circuit diagram for the non-destructive discrimination of $\mathbf{|\psi_{5}\rangle_{m_{1}m_{2}m_{3}m_{4}m_{5}}}$}
	\label{fig:CircuitDiagram}
	\leavevmode
\centering	
\Qcircuit @C=0.8em @R=0.8em {
\lstick{}& \qw &\qw &\qw &\ctrl{1} &\qw &\qw &\ctrl{1} &\qw &\targ &\qw &\qw &\qw &\qw &\ctrl{1}&\qw &\qw &\qw &\qw\\
\lstick{}& \qw &\qw &\ctrl{1}&\qw\qwx &\targ &\qw &\qw\qwx &\ctrl{1} &\qw\qwx &\targ &\qw &\qw &\targ &\qw\qwx &\ctrl{1} &\qw &\qw &\qw\\
\lstick{\ket{\psi_{5}}_{x}}& \qw &\qw &\qw\qwx &\qw\qwx &\ctrl{-1} &\gate{H} &\qw\qwx &\qw\qwx &\qw\qwx &\qw\qwx &\ctrl{1} &\gate{H} &\ctrl{-1} &\qw\qwx &\qw\qwx &\qw &\qw  &\qw &\ \ \ \ |\psi_{5}\rangle_x\\
\lstick{}& \ctrl{1} &\gate{H} &\qw\qwx &\targ\qwx &\qw &\ctrl{1} &\qw\qwx &\qw\qwx &\qw\qwx &\qw\qwx &\qw\qwx &\qw &\qw &\targ\qwx &\qw\qwx &\gate{H} &\ctrl{1} &\qw\\
\lstick{}& \targ\qwx &\qw &\targ\qwx &\ctrl{1} &\qw &\qw\qwx &\qw\qwx &\qw\qwx &\qw\qwx &\qw\qwx &\qw\qwx &\qw &\qw &\qw &\targ\qwx &\qw &\targ\qwx &\qw\\ 
\lstick{\ket{0}}& \qw &\qw &\qw &\qw\qwx &\qw &\qw\qwx &\targ\qwx &\targ\qwx &\qw\qwx &\qw\qwx &\qw\qwx &\qw &\qw &\qw &\qw &\qw &\meter\\
\lstick{\ket{0}}& \qw &\qw &\qw &\qw\qwx &\qw &\qw\qwx &\qw &\qw &\ctrl{-1} &\ctrl{-1} &\qw\qwx &\qw &\qw &\qw &\qw &\gate{\sigma_x} &\meter\\
\lstick{\ket{0}}& \qw &\qw &\qw &\qw\qwx &\qw &\qw\qwx &\qw &\qw &\qw &\qw &\targ\qwx &\qw &\qw &\qw &\qw &\qw &\meter\\
\lstick{\ket{0}}& \qw &\qw &\qw &\qw\qwx &\qw &\targ\qwx &\qw &\qw &\qw &\qw &\qw &\qw &\qw &\qw &\qw &\gate{\sigma_x} &\meter\\
\lstick{\ket{0}}& \qw &\qw &\qw &\targ\qwx &\qw &\qw &\qw &\qw &\qw &\qw &\qw &\qw &\qw &\qw &\qw &\gate{\sigma_x} &\meter
\ \gategroup{1}{19}{5}{19}{.7em}{\}} \gategroup{1}{1}{5}{1}{.7em}{\{}  }  \\
\end{figure}
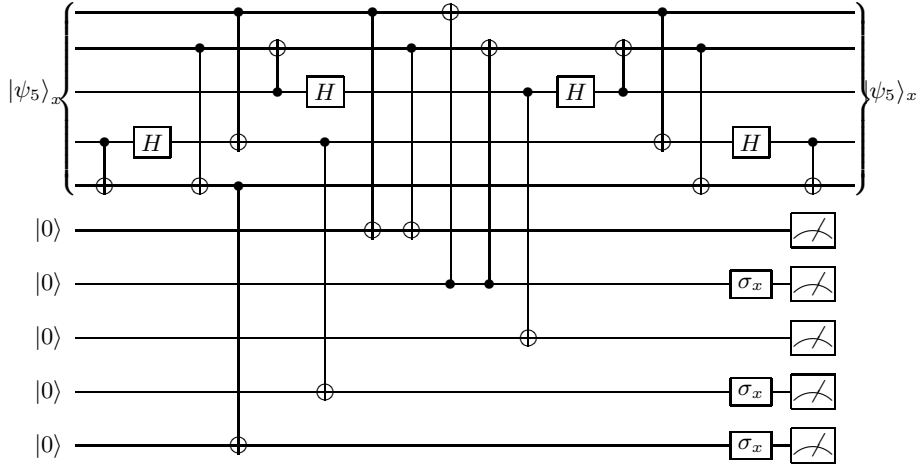

\subsection{Establishment of quantum conversation}
Bob now has all the five qubits of the entangled Brown state. Hence, he can encode his message in the first three qubits by applying unitary operations, as was done earlier by Alice and send the five qubits to Alice in two sets of (4,5) and (1,2,3). On receiving the entangled state Alice applies NDD circuit to obtain the secret information and thus the process continues. Hence, the conversation is established.

\section{Security Analysis}
Suppose Eve wants to attack the quantum channel and eavesdrop Alice's secret message, thereby rendering the protocol worthless. In the following section we show that the protocol is secure against an eavesdropper's attack in ideal conditions i.e., assuming no noises and losses in the quantum channel.

In our protocol, Eve does not have access to all the five qubits at the same time. If she attacks only the encoded particles by Alice, she can gain partial information about the secret message, since the density matrix for Alice's system i.e., $\rho$ =$|001\rangle\langle001|$ + $|010\rangle\langle010|$ + $|100\rangle\langle100|$ + $|111\rangle\langle111|$ is a mixed state, though not maximally mixed, as expected in case of quantum channel with odd number of qubits. Hence, the attack strategy of Eve would be to prepare an ancilla in the initial state $|\eta\rangle$, intercept the qubits 4 and 5 that are being transmitted by Alice to Bob and make the ancilla interact with them using a unitary operation. Next, when the first three encoded qubits are sent to Bob, Eve would capture the qubits and make a joint measurement on the encoded qubits and her ancillas. In this way, Eve might obtain some information about Alice's secret message.

Let the ancilla attached by Eve on the qubits 4 and 5  and the unitary operation performed by her be given by, $|\eta\rangle$  and U respectively. The operations that Eve can perform can be written in the most general form as,
\begin{eqnarray}
 U : |00,\eta\rangle \rightarrow |00,\eta_{00}\rangle + |01,\eta_{01}\rangle + |10,\eta_{02}\rangle + |11,\eta_{03}\rangle\\ \nonumber
|01,\eta\rangle \rightarrow |00,\eta_{10}\rangle + |01,\eta_{11}\rangle + |10,\eta_{12}\rangle + |11,\eta_{13}\rangle\\\nonumber
 |10,\eta\rangle \rightarrow |00,\eta_{20}\rangle + |01,\eta_{21}\rangle + |10,\eta_{22}\rangle + |11,\eta_{23}\rangle\\\nonumber
|11,\eta\rangle \rightarrow |00,\eta_{30}\rangle + |01,\eta_{31}\rangle + |10,\eta_{32}\rangle + |11,\eta_{33}\rangle.
\end{eqnarray}
Here, $|\eta_{ij}\rangle$, (i,j $\in$ \{0,1,2,3\}) are pure ancilla states, uniquely determined by the unitary operator U. After the interaction, the quantum state can be rewritten as,
\begin{eqnarray}
|\varphi\rangle &=& [|\phi_{+}\rangle|1\rangle( |00,\eta_{00}\rangle + |01,\eta_{01}\rangle + |10,\eta_{02}\rangle + |11,\eta_{03}\rangle )\\ \nonumber
& &+ |\psi_{+}\rangle|0\rangle( |00,\eta_{10}\rangle + |01,\eta_{11}\rangle + |10,\eta_{12}\rangle + |11,\eta_{13}\rangle )\\ \nonumber
& &- |\psi_{-}\rangle|0\rangle( |00,\eta_{20}\rangle + |01,\eta_{21}\rangle + |10,\eta_{22}\rangle + |11,\eta_{23}\rangle )\\ \nonumber
& &- |\phi_{-}\rangle|1\rangle( |00,\eta_{30}\rangle + |01,\eta_{31}\rangle + |10,\eta_{32}\rangle + |11,\eta_{33}\rangle )]_{13245E} 
\end{eqnarray}
Since Bob has a choice of measuring in the basis i.e., $BMB_{1}$, $BMB_{2}$ the problem now breaks down into two cases.
In the first case, when Alice and Bob measure in the bases $AMB_{1}$ and $BMB_{1}$ respectively, the state can be written as,
\begin{equation}
|\varphi\rangle_{132E45} = (|\mu_{1}\rangle|\phi_{+}\rangle + |\mu_{2}\rangle|\psi_{+}\rangle + |\mu_{3}\rangle|\phi_{-}\rangle + |\mu_{4}\rangle|\psi_{-}\rangle)_{132E45} 
\end{equation}
where,
\begin{eqnarray}
|\mu_{1}\rangle_{132E} = (|\phi_{+}\rangle|1\rangle|\eta_{00}\rangle + |\psi_{+}\rangle|0\rangle|\eta_{10}\rangle - |\psi_{-}\rangle|0\rangle|\eta_{20}\rangle \\ \nonumber
- |\phi_{-}\rangle|1\rangle|\eta_{30}\rangle) + (|\phi_{+}\rangle|1\rangle|\eta_{03}\rangle + |\psi_{+}\rangle|0\rangle|\eta_{13}\rangle \\ \nonumber
- |\psi_{-}\rangle|0\rangle|\eta_{23}\rangle - |\phi_{-}\rangle|1\rangle|\eta_{33}\rangle ),\\ \nonumber
|\mu_{2}\rangle_{132E} = (|\phi_{+}\rangle|1\rangle|\eta_{01}\rangle + |\psi_{+}\rangle|0\rangle|\eta_{11}\rangle - |\psi_{-}\rangle|0\rangle|\eta_{21}\rangle \\ \nonumber 
- |\phi_{-}\rangle|1\rangle|\eta_{31}\rangle) + (|\phi_{+}\rangle|1\rangle|\eta_{02}\rangle + |\psi_{+}\rangle|0\rangle|\eta_{12}\rangle \\ \nonumber
- |\psi_{-}\rangle|0\rangle|\eta_{22}\rangle - |\phi_{-}\rangle|1\rangle|\eta_{32}\rangle ),\\ \nonumber
|\mu_{3}\rangle_{132E} = (|\phi_{+}\rangle|1\rangle|\eta_{00}\rangle + |\psi_{+}\rangle|0\rangle|\eta_{10}\rangle - |\psi_{-}\rangle|0\rangle|\eta_{20}\rangle \\ \nonumber
- |\phi_{-}\rangle|1\rangle|\eta_{30}\rangle) - (|\phi_{+}\rangle|1\rangle|\eta_{03}\rangle + |\psi_{+}\rangle|0\rangle|\eta_{13}\rangle\\ \nonumber
- |\psi_{-}\rangle|0\rangle|\eta_{23}\rangle - |\phi_{-}\rangle|1\rangle|\eta_{33}\rangle ),\\ \nonumber
|\mu_{4}\rangle_{132E} = (|\phi_{+}\rangle|1\rangle|\eta_{01}\rangle + |\psi_{+}\rangle|0\rangle|\eta_{11}\rangle - |\psi_{-}\rangle|0\rangle|\eta_{21}\rangle \\ \nonumber
- |\phi_{-}\rangle|1\rangle|\eta_{31}\rangle) - (|\phi_{+}\rangle|1\rangle|\eta_{02}\rangle + |\psi_{+}\rangle|0\rangle|\eta_{12}\rangle \\ \nonumber
- |\psi_{-}\rangle|0\rangle|\eta_{22}\rangle - |\phi_{-}\rangle|1\rangle|\eta_{32}\rangle ).
\end{eqnarray}
Comparing equation (8) with (4), we arrive at the following conditions, which should be satisfied in order to avoid introducing error in the measurements:
\begin{eqnarray}
\langle001|\mu_{1}\rangle = \langle100|\mu_{1}\rangle = \langle111|\mu_{1}\rangle = 0,\\ \nonumber
\langle111|\mu_{2}\rangle = \langle001|\mu_{2}\rangle = \langle010|\mu_{2}\rangle = 0,\\ \nonumber
\langle111|\mu_{3}\rangle = \langle100|\mu_{3}\rangle = \langle010|\mu_{3}\rangle = 0,\\ \nonumber
\langle111|\mu_{4}\rangle = \langle001|\mu_{4}\rangle = \langle100|\mu_{4}\rangle = 0.
\end{eqnarray}
Solving the equations (9) and (10), we obtain the following conditions on the ancilla states.
\begin{eqnarray}
|\eta_{00}\rangle = |\eta_{33}\rangle, \ \ \ \ \ \ \ |\eta_{03}\rangle = |\eta_{10}\rangle, \\ \nonumber
|\eta_{22}\rangle = |\eta_{11}\rangle, \ \ \ \ \ \ \ |\eta_{21}\rangle = |\eta_{12}\rangle, \\ \nonumber
|\eta_{20}\rangle = |\eta_{23}\rangle = |\eta_{10}\rangle = |\eta_{13}\rangle = \textbf{0},\\ \nonumber
|\eta_{02}\rangle = |\eta_{01}\rangle = |\eta_{32}\rangle = |\eta_{31}\rangle = \textbf{0}.
\end{eqnarray}
where \textbf{0} is a null ket.

Similarly, the second case is solved where Alice measures in the basis $AMB_{2}$, and Bob measures in the basis $BMB_{2}$ and the following constraints are obtained.
\begin{eqnarray}
|\eta_{00}\rangle = |\eta_{22}\rangle, \ \ \ \ \ \ \ |\eta_{13}\rangle = |\eta_{31}\rangle, \\ \nonumber
|\eta_{11}\rangle = |\eta_{33}\rangle, \ \ \ \ \ \ \ |\eta_{02}\rangle = |\eta_{20}\rangle, \\ \nonumber
|\eta_{01}\rangle = |\eta_{03}\rangle = |\eta_{21}\rangle = |\eta_{23}\rangle = \textbf{0},\\ \nonumber
|\eta_{10}\rangle = |\eta_{12}\rangle = |\eta_{30}\rangle = |\eta_{32}\rangle = \textbf{0}.
\end{eqnarray}

Since both the cases must be satisfied, ancilla qubits must obey both equations (11) and (12) simultaneously. As a result, the entire state reduces to,
 $|\varphi\rangle$ = $|\psi_{5}\rangle|\eta_{00}\rangle_{E}$. From the above equation it is evident that the state $|\varphi\rangle$ is
 reduced to a product of $|\psi_{5}\rangle$ and the ancilla. This implies that Eve cannot gain any information about Alice's secret
 message by measuring the ancillas. Consequently, it can be deduced that the protocol is secure in the conditions, where there are no losses or noises in the quantum channel.

\section{Conclusion}
In conclusion, we have presented a secure ``quantum conversation'' (of classical information) scheme involving dense coding through  highly entangled five qubit states as the  quantum channel. We describe appropriate circuits for the generation of these states which were achieved only through numerical searches in the past. This was carried out using NDD of 32 orthogonal Brown \textit{et al}., states. We also constructed explicit circuit diagrams for the same. Later, we proved that our schemes are secure to eavesdropper's attack in ideal conditions. In future, we wish to make a detailed analysis on the type of states that can be used for ``quantum conversation'' schemes.

\begin{acknowledgments}
The authors acknowledge the NIUS programme undertaken by Homi Bhabha Center for Science Education (HBCSE-TIFR) and Prof. Vijay Singh of
HBCSE for encouragement  and discussions. SJ and SM also acknowledge the warm hospitality of the Indian Institute of Science Education and Research (Kolkata), India, where the work was completed. 

\end{acknowledgments}

\end{document}